\documentclass[12pt,a4paper]{article}

\usepackage[margin=0.7in,top=0.5in,bottom=0.7in]{geometry}
\usepackage{setspace}
\usepackage{graphicx}
\usepackage{wrapfig}
\usepackage{bm}
\usepackage{color}

\usepackage[parfill]{parskip} 

\usepackage{times}

\newcommand{\eq}{\begin{equation}}
\newcommand{\eeq}{\end{equation}}
\newcommand{\eqa}{\begin{eqnarray}}
\newcommand{\eeqa}{\end{eqnarray}}

\newcommand{\dee}{{\rm d}}

\begin{document}

\title{Interstitial-mediated dislocation climb and the weakening of particle-reinforced alloys under irradiation}
\author{DH Thompson${}^1$, E Tarleton${}^1$, SG Roberts${}^1$ and SP Fitzgerald${}^{2}\footnote{S.P.Fitzgerald@leeds.ac.uk}$\vspace{5mm} \\ 
${}^1$Department of Materials, University of Oxford, OX1 3PH\\
${}^2$Department of Applied Mathematics, University of Leeds, LS2 9JT}

\date{\today}

\maketitle
\begin{doublespacing}

Dislocations can climb out of their glide plane by absorbing (or emitting) point defects (vacancies and self-interstitial atoms (SIAs)). In contrast with conservative glide motion, climb relies on the point defects' thermal diffusion and hence operates on much longer timescales, leading to some forms of creep. Whilst equilibrium point defect concentrations allow dislocations to climb to relieve non-glide stresses, point defect supersaturations also lead to osmotic forces, driving dislocation motion even in the absence of external stresses. Self-interstitial atoms typically have significantly higher formation energies than vacancies, so their contribution to climb is usually ignored. However, under irradiation conditions, both types of defect are athermally created in equal numbers. In this letter, we use simple thermodynamic arguments to show that the contribution of interstitials cannot be neglected in irradiated materials, and that the osmotic force they induce on dislocations is many orders of magnitude larger than that caused by vacancies. This explains why the prismatic dislocation loops observed by {\it in situ} transmission electron microscope irradiations are more often of interstitial rather than vacancy character. Using discrete dislocation dynamics simulations, we investigate the effect on dislocation-obstacle interactions, and find reductions in the depinning time of many orders of magnitude. This has important consequences for the strength of particle-reinforced alloys under irradiation. 

At finite temperature $T$, crystalline materials contain vacancies and SIAs with thermal equilibrium concentrations given approximately by $c_{\rm therm}=\exp\left( -E_{\rm form}/k_{\rm B}T\right)$, where $E_{\rm form}$ is the formation energy of the vacancy or SIA, and $k_{\rm B}$ is Boltzmann's constant \footnote{This is the absolute concentration, $10^{-6}$ times the parts per million value.}. This reflects the probability of a lattice site being ``occupied'' by a vacancy or SIA \cite{hirth1982}. Irradiation generates microstructure by displacing atoms from their lattice sites, creating Frenkel pairs of vacancy and SIA point defects \cite{was2016} in addition to the thermal equilibrium concentrations already present, so the crystal can lower its energy by removing the excess defects via recombination, annihilation at sinks such as grain boundaries or dislocations, and agglomeration into prismatic dislocation loops and voids. This microstructure causes the swelling, hardening, and embrittlement whose mitigation is a central focus of nuclear materials science.

Over the past 60 years, a large amount of work has been devoted to modelling the diffusion and clustering of point defects into loops and voids, in particular the calculation of bias factors quantifying the preferential absorption of SIAs over vacancies at sinks, thus leading to void swelling (see e.g. refs.\cite{brailsford1972,wolfer2007} among many others). Less attention has been focussed on the effects out-of-equilibrium defect concentrations have on the dynamics of the existing dislocation network, and hence on mechanical properties. Mansur \cite{mansur1979} developed a model of irradiation creep by climb-enabled glide due to stress-induced preferred absorption of SIAs at dislocations. Raabe \cite{Raabe1998} explored all the possible origins for forces on dislocations, and explained how to incorporate osmotic forces due to a vacancy supersaturations into discrete dislocation dynamics (DDD). Mordehai \emph{et al} \cite{mordehai2008} developed a practical implementation, and Bako \emph{et al} \cite{bako2011} used this to model prismatic loop coarsening by climb. Danas \cite{danas2013} included climb and climb-assisted glide in large scale ``2.5 dimensional'' DDD simulations, where some 3D phenomena are taken into account in an efficient 2D model, while Gao {\it et al} \cite{gao2017} applied a hybrid simulation to creep in Ni-based superalloys. With the exception of ref. \cite{mansur1979}, these works neglect the contribution to climb made by SIAs. This is entirely reasonable when thermal concentrations of point defects are considered, since the formation energy of SIAs is typically twice that of vacancies in metals \cite{derlet2007}, and so the thermal equilibrium concentration of SIAs will be far smaller than that of vacancies. Under irradiation, however, superthermal concentrations of both types of point defects are generated in equal numbers, and the tiny thermal concentration for SIAs means the \emph{relative} concentration, and hence the osmotic force exerted, is much larger than that for superthermal vacancies. Fu {\it et al} \cite{fu2017} carried out molecular dynamics simulations of irradiation cascades near edge dislocations in tungsten, and found strong effects on edge dislocation structures. 

%

In this letter, we modify Mordehai \emph{et al}'s climb model to include SIAs in addition to vacancies (see Methods). Considering the SIA osmotic force effects on prismatic loop growth (or shrinkage) in the presence of supersaturated vacancy and SIA concentrations, as a function of temperature, we suggest an experimental method to determine effective steady-state supersaturations in {\it in situ} transmission electron microscope (TEM) irradiations. We then incorporate the SIAs into a climb velocity rule for DDD simulations, with which we simulate dislocation shear loop depinning from hard sphere obstacles. This serves as a simple model for the strengthening mechanism employed in oxide-dispersion-strengthened (ODS) steels. We find that the radiation-induced SIAs significantly enhance the climb mobility of dislocations, leading to fast depinning at far lower temperatures than for thermal vacancy climb. We discuss the results and implications in the context of available literature data on \emph{in situ} TEM of prismatic loop evolution \cite{xu2009}, and yield strength of and irradiation creep of some nuclear steels \cite{toloczko2001,toloczko2004,elgenk2005,kim2003}.

The mechanical force $\bm{F}$ acting on a dislocation with Burgers vector $\bm{b}$ and line direction $\bm{\hat l}$, due to a stress $\bm{\sigma}$, is given by the Peach Koehler formula $\bm{F} = \left(\bm{\sigma\cdot b}\right)\times \bm{\hat l}$. In the absence of point defects, non-screw dislocations are confined to the glide plane containing $\bm{b}$ and $\bm{\hat l}$, and the component of $\bm{F}$ acting normal to this plane cannot be relieved. This, and hence climb, can only happen by the absorption (or emission, under extremely high climb stresses) of point defects, and is hence generally limited by the availability and mobility of these defects. Under normal conditions, this happens much more slowly than mechanical motion within the glide plane. 

\begin{figure}
\begin{center}
\includegraphics[width=0.45\textwidth]{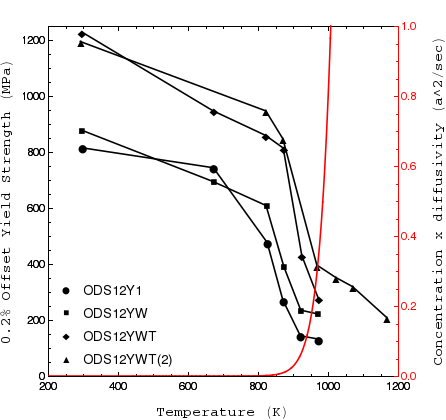}
\includegraphics[width=0.45\textwidth]{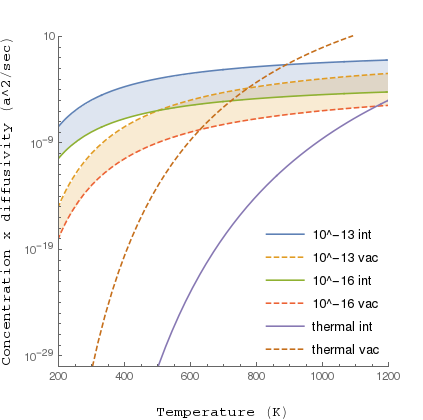}
\caption{Left: yield strength of several ODS steels vs. temperature \cite{elgenk2005,kim2003}; temperature dependence of thermal vacancy climb velocity in Fe (units (lattice parameter $a$)${}^2$/second. An additional factor of $|\bm{b}|$ enters in the denominator to make this a velocity). The steep fall in yield strength coincides with the sharp rise in climb velocity. Right: temperature dependence of climb velocity due to thermal concentrations of vacancies (dashed lines) and SIAs (solid lines), plus superthermal concentrations of $10^{-16}-10^{-13}$ shown in bands. Supersaturations, particularly of SIAs, enhance climb by many orders of magnitude at lower temperatures.}\label{fig:2}
\end{center}
\end{figure}

The climb velocity (see ref.\cite{mordehai2008} and Methods) depends on temperature $T$ via the product of the point defect concentration $c_{\rm therm}$ and diffusivity $D$: $c_{\rm therm}(T)=\exp(-E_{\rm form}/k_{\rm B}T);\; D(T)=\; D_0\exp(-E_{\rm migr}/k_{\rm B} T)$, where $E_{\rm form}$ and $E_{\rm migr}$ are formation and migration energies respectively, and $D_0$ is the diffusion prefactor typically taken to be $10^{13}$Hz \cite{fu2005}. In Fe, $E_{\rm form}\approx$ 2eV for vacancies, and 4eV for interstitials, whilst $E_{\rm migr}\approx$ 0.6eV for vacancies and 0.3eV for interstitials \cite{derlet2007}. Their larger formation energy means thermal interstitials remain negligible over the relevant temperature range, as shown in Fig.\ref{fig:2}, (right), but the climb velocity due to thermal vacancies in Fe increases by some 30 orders of magnitude as temperature increases from 300 to 1000K.

Fig.\ref{fig:2} (left) shows the 0.2\% offset yield strength of several (unirradiated) oxide-dispersion-strengthened (ODS) steel, 12YWT \cite{elgenk2005,kim2003}. This and similar alloys employ hard ceramic particles, dispersed throughout the matrix, to pin dislocations, resulting in high room temperature yield strength of almost 1200MPa. By 1000K this strength falls to less than a third of its room temperature value, and the high temperature strength is comparable to that of non-reinforced alloys, implying that the particle-based strengthening mechanism is no longer operative. Also shown on Fig.\ref{fig:2} (left) is the temperature-dependent part of the dislocation climb velocity due to thermal vacancies, which rises to one lattice parameter per second at 1000K (an additional factor of $|\bm{b}|$ appears in the dislocation velocity expression, see formula given in Methods). This increased vacancy concentration and mobility allows pinned dislocations to climb around the oxide particles above 800K, negating their strengthening effect. Screw dislocations can overcome obstacles by cross-slip, but this process has a low activation barrier in bcc crystals, and cannot explain the observed temperature dependence. Another possibility is the ``edge cross-slip'' process suggested by Humphreys {\it et al} \cite{humphreys1970}. However, this involves the generation of a screw dipole around the obstacles, whose sides must cross-slip in the same direction under opposite Peach-Koehler forces. Thus we conclude that the dominant depinning mechanism is vacancy-mediated dislocation climb. In what follows, we show that the presence of non-equilibrium supersaturations of SIAs induced by irradiation can lead to climb-enabled depinning at temperatures much lower than 1000K.  

When $c_{\rm therm}$ is replaced by a superthermal vacancy or SIA concentration $c>c_{\rm therm}=\exp(-E_{\rm form}/k_{\rm B}T)$ is present, as is the case under irradiation, an additional ``osmotic'' climb force on the dislocation arises, due to the nonequilibrium chemical potential of the defects: $F_{\rm os} \propto k_{\rm B}T\left(1-c/c_{\rm therm}\right) =  k_{\rm B}T\left(1-c\exp\left[ +E_{\rm form}/k_{\rm B}T\right]\right)$ \cite{mordehai2008}. This is the thermodynamic driving force acting to remove the supersaturated defects from solution, and causes dislocation climb even in the absence of a mechanical stress. Crucially, for a given superthermal concentration, this force is much larger for interstitials than for vacancies, since their formation energy is doubled and the energy dependence is exponential, and hence there is a far greater energetic advantage to removing them from solution. As temperature increases for a fixed superthermal concentration $c$, the osmotic force reduces as $c_{\rm therm}$ increases, eventually falling to zero when $c_{\rm therm}$ reaches $c$. 

\begin{figure}
\begin{center}
\includegraphics[width=0.45\textwidth]{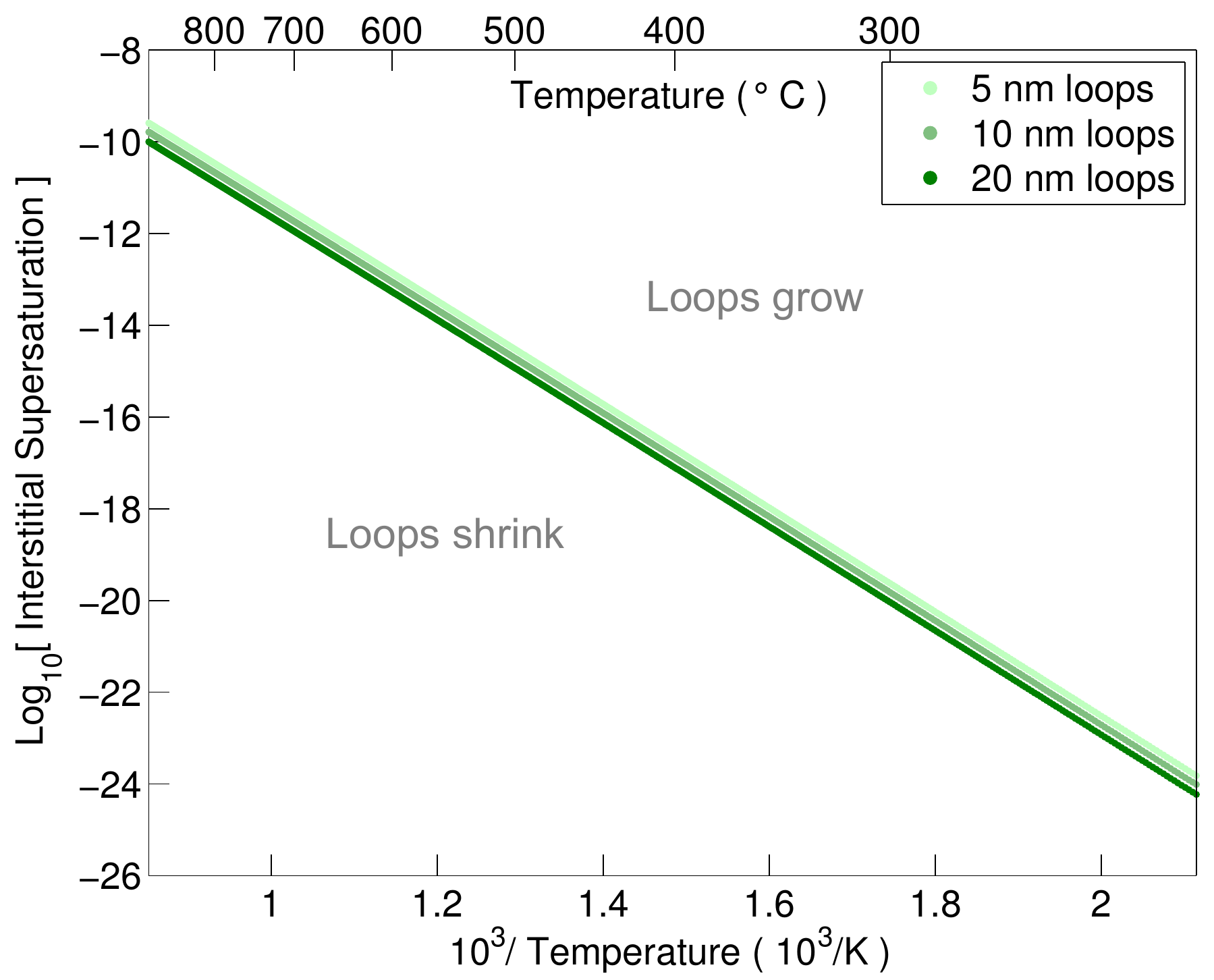}
\includegraphics[width=0.45\textwidth]{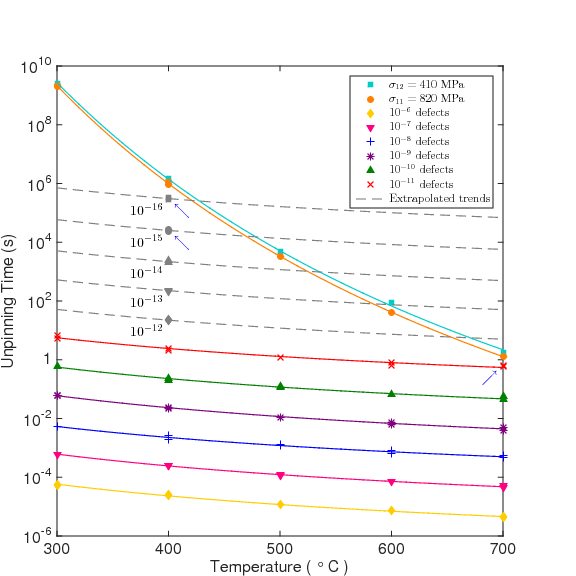}
\caption{Left: transition temperature where interstitial osmotic force and line tension self force balance for prismatic loops. Right: depinning time vs. temperature for applied stress only (thermal vacancy climb, upper two curves) and osmotic forces due to supersaturated SIAs.}\label{fig:3}
\end{center}
\end{figure}

Fig.\ref{fig:2} (right) compares $c(T)D(T)$ for thermal equilibrium defects with imposed concentrations of $10^{-16}$-$10^{-13}$, or $10^{-10}$-$10^{-7}$ ppm. For these levels of supersaturation, the SIA contribution to dislocation climb at 400K is comparable to that of thermal vacancies at 700-800K. The level of supersaturation encountered in a real irradiated material is difficult to determine, and it depends on the material and its thermal and mechanical history as well as irradiation conditions. The non-equilibrium steady state supersaturation we consider here is an {\it effective} value, assumed to be reached when the number of defects injected by cascades per second is balanced by the (concentration-dependent) rate of annihilation by recombination and absorption at sinks. For the case of {\it in situ} ion-beam TEM irradiations, a lower bound on the effective supersaturation can be found by considering the prismatic interstitial dislocation loops formed under these conditions. Such a loop of radius $R$ experiences an elastic climb self-force \cite{gavazza1976} acting to shrink itself of magnitude 
 $F_{\rm self} = -\frac{\mu b^2}{4\pi (1-\nu)}\frac1{R}\left(\ln\left(\frac{8R}{r_c}\right)-1+\frac{3-2\nu}{4(1-\nu)}\right).$ This can be relieved by vacancy climb, and acts in the opposite direction to any interstitial osmotic force acting to grow the loop. This provides a physical explanation for the dominant observation of interstitial loops in the TEM: for equal vacancy and SIA supersaturations, the osmotic force acting to push the SIAs out of solution into prismatic loops is far greater than that for vacancies, because their equilibrium thermal concentration is far lower. This outweighs the increased loss of highly mobile SIAs and clusters to the surface of the TEM foil. The vacancies can remain in solution up to far higher concentrations due to their lower formation energy. Furthermore, vacancy loops' self stress can be relieved by SIA-mediated climb, shrinking or removing vacancy loops, which reinforces the mechanism. As temperature increases, the thermal equilibrium interstitial concentration increases, so the osmotic force due to excess SIAs reduces, and the vacancy climb mobility increases. Thus a transition temperature will be reached where, for a fixed SIA supersaturation, interstitial loops of a given size will stop growing when their vacancy-mediated, self-force-induced climb acting to shrink the loop balances the SIA osmotic force acting to grow it. Fig.\ref{fig:3} (left) shows the predicted transition temperature vs. supersaturation for 5, 10 and 20nm diameter SIA loops. Xu {\it et al} \cite{xu2009} observed SIA loop growth at 460${}^{\circ}$C, pointing to supersaturations of at least $10^{-16}$, but {\it in situ} irradiations at higher temperatures are required to identify the transition we predict. 

\begin{figure}
\begin{center}
\includegraphics[width=0.31\textwidth]{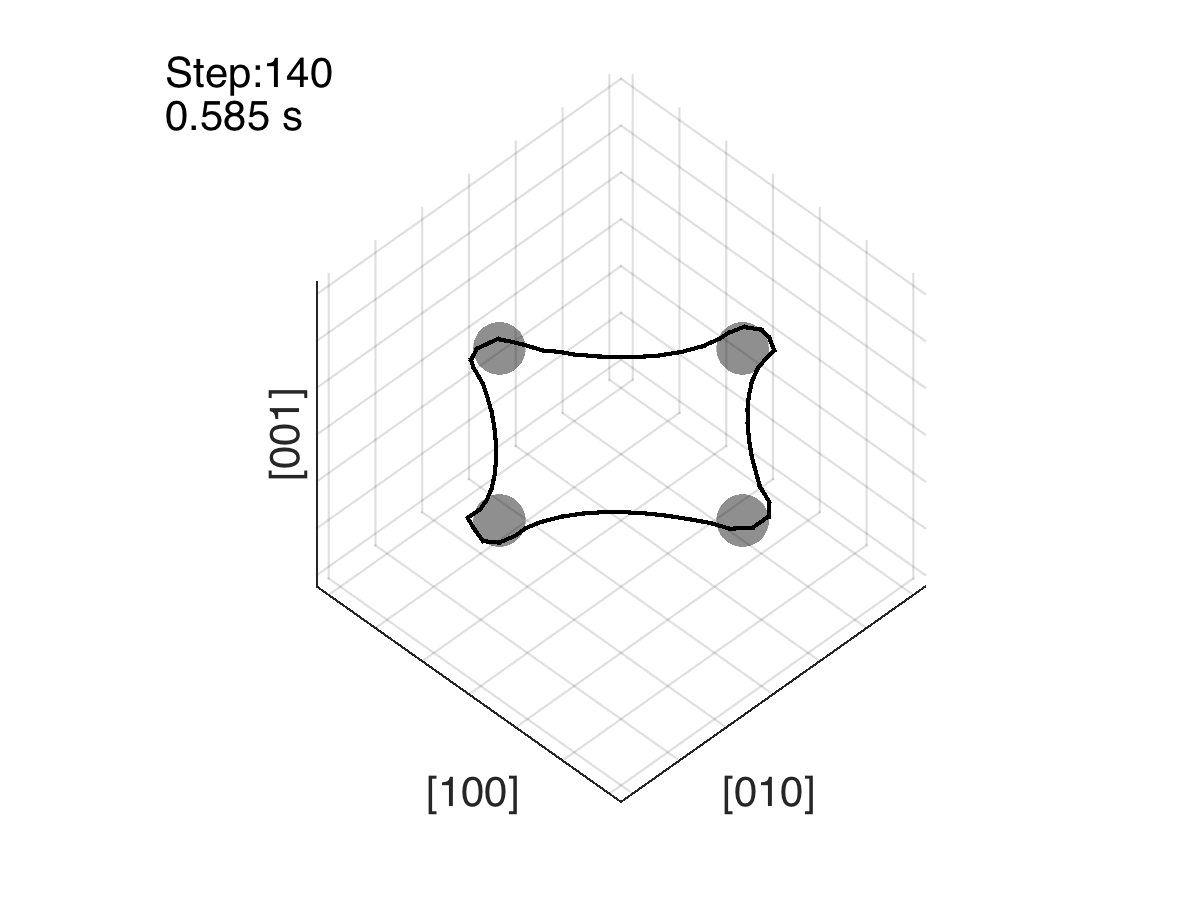}
\includegraphics[width=0.31\textwidth]{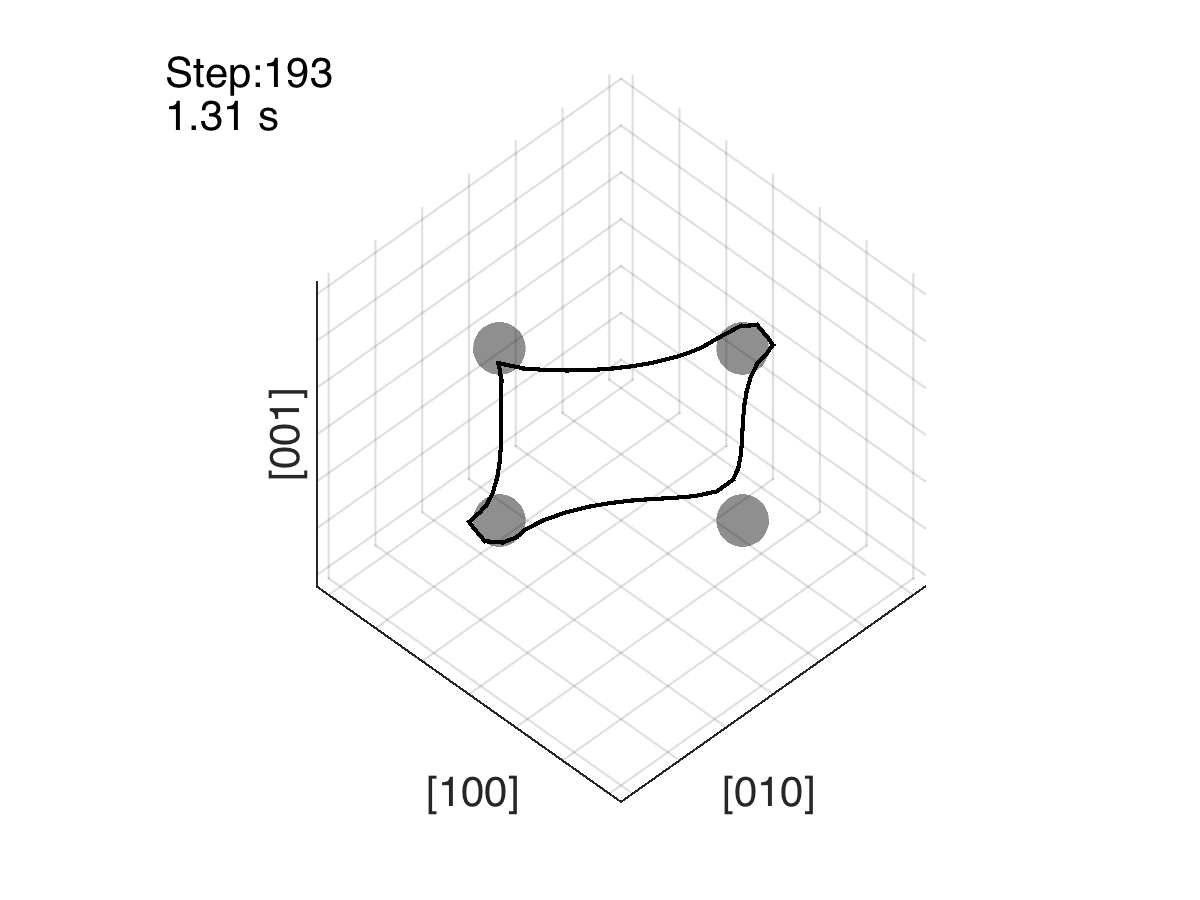}
\includegraphics[width=0.31\textwidth]{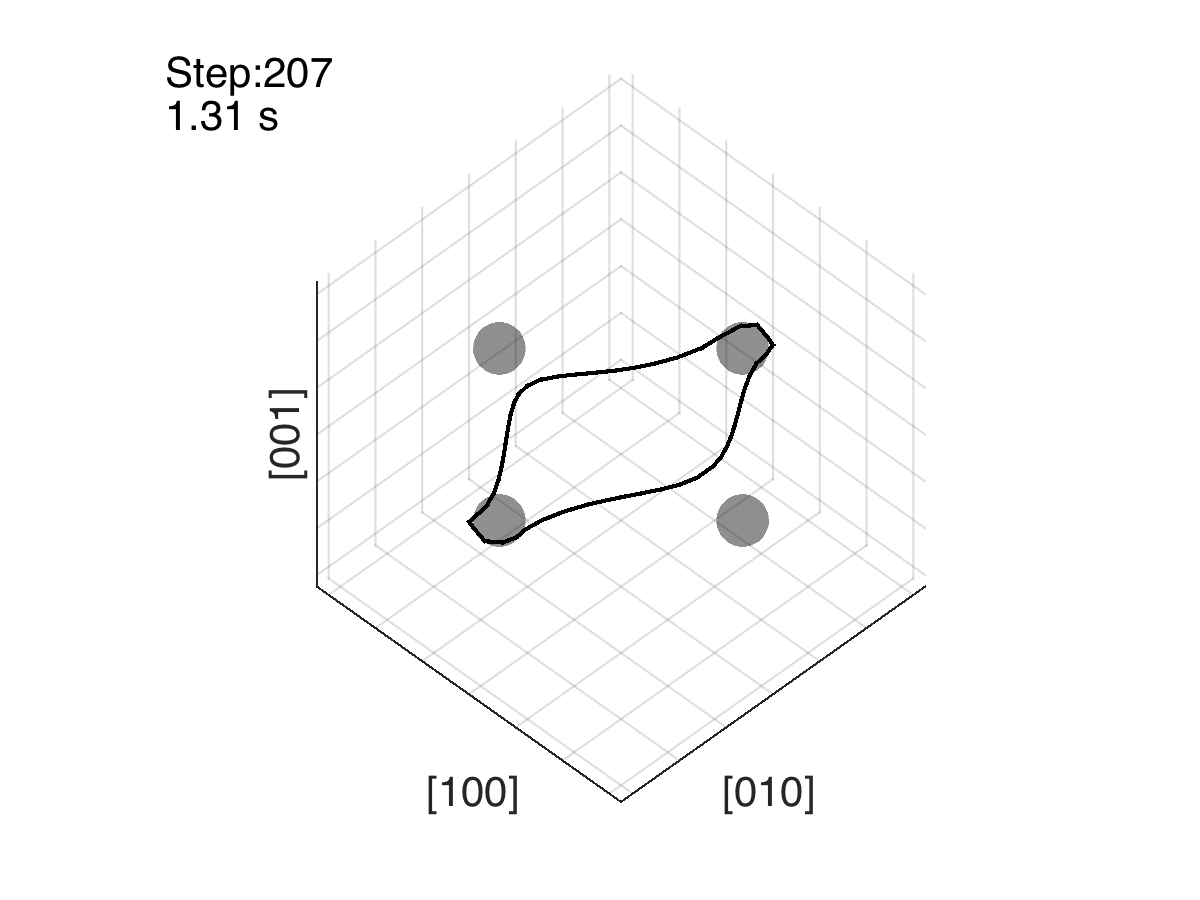}
\caption{Discrete dislocation dynamics simulation setup. The pure screw parts of the shear loop meet the obstacles at the upper right and lower left. }\label{fig:4}
\end{center}
\end{figure}

In DDD simulations, climb can be introduced by allowing a small mobility in response to the climb component of the Peach-Koehler force on each dislocation segment (typically $10^{-4}$ of the glide mobility \cite{bulatov2006,arsenlis2007}). Even with the major enhancement provided by the point defect supersaturations, climb is slower than glide, and to improve simulation efficiency we implemented a dual time step procedure, where dislocations glide until pinned, then climb until an appreciable glide force arises. To model depinning in ODS alloys, we placed a shear loop in the plane containing an array of four impenetrable spherical inclusions in an infinite elastic solid. The loop shrinks by glide under its own self-force until it becomes pinned on the inclusions, after which the non-screw segments begin to climb in the presence of superthermal point defect concentrations (fig.\ref{fig:4}). Fig.\ref{fig:3} (right) shows the time taken for the loop to unpin from the inclusions vs. temperature for supersaturations of $10^{-6}-10^{-16}$ (the lower supersaturations are extrapolated from the clear functional dependence, see Methods), together with thermal depinning times under assisting applied loads of 410 and 820MPa. The thermal depinning time is strongly temperature-dependent, falling from over 30 years at 300${}^{\circ}$C to around 1 second at 700${}^{\circ}$C. This is due to the nonlinearity in temperature of the Arrhenius function ${\rm e}^{-E/k_{\rm B}T}$, and is consistent with the precipitous fall in strength the unirradiated ODS steels exhibit at high temperature (fig.\ref{fig:2}, \cite{elgenk2005,kim2003}). Equivalent depinning times occur at just 300${}^{\circ}$C with supersaturations of $10^{-11}$, with commensurate reductions in depinning time for intermediate supersaturations. This is the central result of this letter: {\bf irradiation-induced supersaturations of point defects, in particular SIAs, increase the climb rate at low and intermediate temperatures to levels associated with thermal climb at much higher temperatures.} Dispersed particles cease to be effective barriers once the climb rate becomes high enough to allow dislocations to quickly climb around them. Without irradiation, this occurs above around 800K in the ODS alloys we consider. Irradiation means this loss of strength occurs at much lower temperatures. 

Quantitative experimental verification of the SIA-enhanced climb predicted here is challenging, chiefly because it is an inherently out-of-equilibrium phenomenon. {\it Ex situ} post-irradiation mechanical testing cannot hope to capture the effects, which rely on a steady-state superthermal defect concentration being maintained. In-reactor creep tests are time-consuming and expensive, and in the past have been restricted to candidate reactor steels such as HT-9 \cite{toloczko2001} and ODS alloys such as those mentioned above \cite{toloczko2004}. Such materials are necessarily complicated, and it is difficult or impossible to extract precise information on individual microstructural processes. Large-grained pure metal samples with oxide particle reinforcement would be more tractable, but such model materials would of course have no structural application themselves. 

Toloczko {\it et al} \cite{toloczko2001} have compared thermal with irradiation creep in HT-9, and report greatly increased creep rates at low stresses under irradiation below 500${}^{\circ}$C, with thermal creep catching up above 600${}^{\circ}$C. This is consistent with the discussion above -- with an effective excess SIA concentration of order $10^{-13}$, Fig.\ref{fig:2} (right) shows the effect of the SIA osmotic pressure at 700K enhancing climb to levels not reached until 900K in the absence of irradiation. However, it should be stressed that a quantitative comparison is not possible: the creep tests cited above use pressurized tubes to impose a uniform hydrostatic load up to around 200MPa -- only a fraction of the unirradiated yield stress, which is of order 1 GPa for the ODS alloys we consider. Whilst excess point defects can and clearly do increase creep rates under these circumstances, the process is isotropic, and is likely due to the dislocation bias effects (prismatic loop growth, void formation etc) cited above, rather than the assisted depinning of dislocations from obstacles we describe here. Under the large directional shear loads of a tensile test, dislocation depinning plays a central role. 

One possibility would be to exploit recent advances in micromechanical testing, and to use an ion or proton beam to generate continuous displacement damage whilst applying a tensile load to a micron-scale sample, similar to the helium irradiations carried out in Ref. \cite{chen2013}.

{\bf The mechanism we describe significantly reduces the efficacy of particle reinforcement under irradiation, resulting in severe loss of strength at temperatures far lower than mechanical tests performed on unirradiated samples suggest.} {\it In situ} tensile tests are required to investigate this phenomenon experimentally. (We note that ODS particles can play other useful roles besides pinning dislocations, in particular providing sinks for radiation-induced point defects and gas atoms).

\begin{small}

{\bf Methods}

{\it Bulk diffusion model --} (see ref.\cite{mordehai2008} for a thorough discussion). The dislocation climb velocity is computed from the point defect flux to the core, which is given by the steady-state coupled reaction-diffusion equations for the vacancy and SIA concentrations $c_v(\bm{r}) ,c_i(\bm{r})$:
\eqa
\frac{\partial c_v}{\partial t} & = & D_v\nabla^2 c_v + K_v - Ac_v c_i = 0\nonumber\\
\frac{\partial c_i}{\partial t} & = & D_i\nabla^2 c_i + K_i - Ac_i c_v = 0,\nonumber
\eeqa where $K_{v,i}$ are point defect creation rates, $A$ is the recombination rate and the $D$s are the diffusion constants. Neglecting the elastic interactions between the point defects and the dislocation (which is a good approximation in the steady state \cite{danas2013,ham1959}) simplifies the problem to a cylindrically-symmetric one. Assuming further that $K_v=K_i$, the function $u(r)\equiv D_vc_v(r)-D_ic_i(r)$ satisfies the Laplace equation. This is solved subject to boundary conditions $c_{v,i}\to c^{\infty}_{v,i}$ as $r\to \infty$, and $c_{v,i}(r_0)=\exp\left(-\left(E_{v,i}^{\rm form}\pm F_{\rm cl}\Omega/|\bm{b}|\right)/k_{\rm B}T\right)$ at the dislocation core radius $r_0$, which is assumed to act as a perfect sink. $F_{\rm cl}$ is the climb component of the Peach-Koehler force. The Burgers' vector is $\bm{b}$ and the atomic volume is $\Omega$. This condition imposes {\it local} equilibrium at the core of the climbing dislocation \cite{mordehai2008}, and reflects the change in the effective formation energy of a point defect due to the potential work done by (or against) the climb stress on absorption (or emission) of a defect. The $\pm$ sign is chosen depending on the sign of the climb stress. The dislocation climb velocity is then proportional to the difference between the vacancy and SIA flux to the core, $\oint_{r=r_0}\nabla u\cdot \bm{\hat n}r_0\dee \phi$:
\eqa
v_{\rm cl} & = & \frac{2\pi\Omega}{|\bm{b}|^2k_{\rm B}T\ln(r_{\infty}/r_0)}D_0{\rm e}^{-(E_{\rm form}^v + E_{\rm migr}^v)/k_{\rm B}T}\left(F_{\rm cl} \pm\frac{|\bm{b}|k_{\rm B}T}{\Omega}\left(1-c_v^{\infty}{\rm e}^{+(E_{\rm form}^v/k_{\rm B}T}\right)  \right) \nonumber\\
& + & v\rightarrow i;\;\; \pm\rightarrow\mp, 
\eeqa where the second line is the interstitial contribution, obtained by replacing $v$ with $i$ in the first and reversing the appropriate signs, and we have expanded in powers of $F_{\rm cl}\Omega/|\bm{b}|k_{\rm B}T$ keeping only the linear term. The mechanical and osmotic forces thus separate into the two terms in parentheses. $r_{\infty}$ is an outer cut-off, typically taken to be the grain size. The ratio of the two lengthscales, entering logarithmically, can be adjusted to fit experimental or numerical data. Note that the explicit creation and recombination rates cancel at this level of approximation, and all information about the irradiation is contained in the effective supersaturation levels, $c_{v,i}^{\infty}$. The depinning time is clearly inversely proportional to $v_{\rm cl}$, as our simulations confirm (Fig.\ref{fig:3}, right). Various configurations of Burgers vector and slip system were investigated; the differences between them are invisible on the log scales of Fig.\ref{fig:3}.

{\it DDD simulations --} We modified a version of the \texttt{DDLab/ParaDiS} code described in detail in refs.\cite{bulatov2006,arsenlis2007}. The shear modulus was 82GPa, the Poisson ratio was 0.305, and the lattice parameter was 2.87\AA. Vacancy formation and migration energies were 2 and 0.6eV respectively; the values for interstitials were 4 and 0.3eV, corresponding to iron.

\end{small}

\end{doublespacing}


\end{document}